\documentclass[fleqn,twoside,twocolumn,nofootinbib,showkeys]{revtex4} 
\usepackage[sec,nocpr]{ujp} 

\begin{document}
\title[Effective Potential of Electron-Electron Interaction]
{EFFECTIVE POTENTIAL\\ OF ELECTRON-ELECTRON INTERACTION\\ IN THE
SEMIINFINITE
ELECTRON GAS\\ WITH REGARD FOR THE LOCAL-FIELD CORRECTION}%
\author{B.M.~Markovych}
\affiliation{National University ``L'vivs'ka Politekhnika''}
\address{12, S. Bandera Str., Lviv 79013,
Ukraine}
\email{bogdan_markovych@yahoo.com, ivanzadv@yahoo.com}
\author{I.M.~Zadvorniak}%
\affiliation{National University ``L'vivs'ka Politekhnika''}%
\address{12, S. Bandera Str., Lviv 79013,
Ukraine}%
\email{ivanzadv@yahoo.com} \udk{530.145} 
\razd{\secix}

\autorcol{B.M.\hspace*{0.7mm}Markovych,
I.M.\hspace*{0.7mm}Zadvorniak}

\newcommand{\dd}{\mathrm{d}}
\newcommand{\ee}{\mathrm{e}}
\newcommand{\ii}{\mathrm{i}}
\newcommand{\spc}{\!\!\!\!}
\newcommand{\Sp}{\mathrm{Sp}\,}

\setcounter{page}{1107}%

\begin{abstract}
The effective potential of electron--electron interaction and the
two-particle \textquotedblleft density--density\textquotedblright\
correlation function have been calculated for a simple semiinfinite
metal making allowance for the local-field correction.\,\,The
influences of a flat interface and various models of local-field
correction on the results of calculations are analyzed.
\end{abstract}

\maketitle

\section{Introduction}\vspace*{1mm}

The modern quantum-mechanical statistical theory of Fermi systems
with an interface is still far from being completed.\,\,The urgency
in the theoretical description of such systems can hardly be
overestimated owing to the importance of processes that occur in the
presence of an interface and a rapid development of experimental
methods aimed at surface researches.

The most popular theoretical method to study such systems is the
density functional one \cite{Dreizler}, which was created on the
basis of the well-known Thomas--Fermi approximation firstly
developed for atoms.\,\,By its nature, the density functional method
is a one-particle approach, so it cannot consider many-body
correlation effects correctly.\,\,Therefore, the energy functionals
of systems with an interface are most often considered in the local
density approximation; namely, the expressions known from the theory
of uniform systems are taken for calculations, but the distribution
of the electron density $n(\mathbf{r})$ is substituted for the
electron concentration $n$.\,\,This approach is debatable
\cite{Sarry}, because the interface introduces not only
quantitative, but also qualitative changes in various parameters of
the electron system (e.g., the image forces emerge and so on), which
cannot be taken into account in principle by the density functional
theory.

In the cycle of works
\cite{JPS2003:1,JPS2003:2,CMP2003,KM2006CMP,KMK2006ICMP,KM2008CMP,KM2009AIP},
an attempt was made to develop a consistent quantum-mechanical
statistical theory for a simple metal with the interface
\textquotedblleft me\-tal--va\-cu\-um\textquotedblright.\,\,In
particular, it was shown that the thermodynamic potential and the
structural distribution functions of electrons in a semiinfinite
metal are expressed in terms of the effective electron-electron
interaction potential.\,\,This work logically continues this cycle
of works.\,\,It was aimed at studying the influence of various
approximations for the local-field correction on the two-particle
correlation function \textquotedblleft
den\-sity--density\textquotedblright\ and the effective potential of
electron-electron interaction.

The effective interaction between charged particles in spatially
confined systems attracts the attention of researchers for a long
time.\,\,In particular, it was studied in works
\cite{Sidyakin,Gabov1,Beck1,Beck2,Gabov2,Vecchio,Bechstedt,Fano,Fratesi}.
In work \cite{Sidyakin}, under certain approximations, the
polarization part of the energy of interaction between a motionless
point charge and a semiinfinite metal was calculated.\,\,Using
similar approximations, the dielectric function of a semiinfinite
metal and the effective potential were calculated in work
\cite{Bechstedt}.\,\,In works \cite{Gabov1,Gabov2} in the framework
of the effective interaction potential approximation, the
asymptotics of Friedel oscillations at large distances between
charges at the interface was obtained, and their dependence on the
Fermi surface shape in metals were studied.

In works \cite{Beck1,Beck2,Vecchio,Fano}, the screening of a charged
impurity near the metal surface was studied in the random-phase
approximation.\,\,In particular, the dependences of the
electrostatic potential of this impurity on the distance in the
surface plane and the distance to the surface were calculated in the
quasiclassical case and the Thomas--Fermi approximation, and Friedel
oscillations were revealed.

In order to eliminate shortcomings inherent to the density
functional method, attempts were made to use the GW approach
\cite{Aryasetiawan} with the density functional method.\,\,In
particular, in work \cite{Fratesi}, the effective potential of
electron-electron interaction was calculated with the use of the
local density approximation for the exchange-correlation potential.

\section{Model}

Consider a semiinfinite simple metal in the framework of the jellium model, i.e.
when the ionic subsystem of a metal is represented by a uniformly distributed
positive charge confined by the interface plane $z=$ $=-d$, with the density%
\begin{equation}
n_{+}(x,y,z)\equiv n_{+}(z)=n_{\mathrm{bulk}}\,\theta(-d-z),
\end{equation}
where\vspace*{-2mm}
\[
  \theta(x)=\left\{\!\!
  \begin{array}{cc}
    1,\quad x>0, \\
    0,\quad x<0 \\
  \end{array}\right.
 \]
is the Heaviside function, $n_{\mathrm{bulk}}$ is the electron
concentration, and $d>0$ is a parameter determined self-consistently
from the electroneutrality condition
\begin{equation}
\int\limits_{-\infty}^{+\infty}\!\mathrm{d}z\left(  n(z)-n_{+}(z)\right)
=0,\label{nejtr}%
\end{equation}
where $n(z)$ is the electron density distribution.\,\,Let the ionic
subsystem form a surface potential for electrons in the metal, which
does not allow them to escape.\,\,This surface potential is
simulated by the potential wall
\begin{equation} \label{GrindEQ__1_}
 V\left(z\right)=
 \left\{\!\!
  \begin{array}{ll}
  \infty,& z\geqslant0, \\
  0,    & z<0.
 \end{array}\right.
 \end{equation}
This model of potential physically correctly corresponds to a real
situation and allows analytical solutions to be obtained for the
Schr\"{o}dinger equation\vspace*{-1mm}
\[
\left[  -\frac{\hbar^{2}}{2m}\Delta+V(z)\right]  \Psi_{\mathbf{p},\alpha
}(\mathbf{r})=E_{\alpha}(\mathbf{p})\Psi_{\mathbf{p},\alpha}(\mathbf{r}),
\]
where $m$ is the electron mass, $\hbar\mathbf{p}$ the
two-dimensional vector of electron momentum in the plane parallel to
the interface, and $\alpha$ the quantum number associated with the
electron motion normally to the interface.\,\,Since the electron
moves freely in parallel to the interface, the wave function and the
corresponding energy of the electron can be written as
follows:\vspace*{-1mm}
\[
\Psi_{\mathbf{p},\alpha}(\mathbf{r})=\frac{1}{\sqrt{S}}\mathrm{e}%
^{\mathrm{i}\mathbf{p}\mathbf{r}_{||}}\varphi_{\alpha}(z),\quad E_{\alpha
}(\mathbf{p})=\frac{\hbar^{2}(p^{2}+\alpha^{2})}{2m}.
\]
Here, $\mathbf{r}=(\mathbf{r}_{||},z)$ is the radius vector of the
electron, $S$ is the interface area, and\vspace*{-1mm}
\[
\varphi_{\alpha}(z)=\frac{2}{\sqrt{L}}\sin(\alpha
z)\theta(-z),\;\alpha =\frac{2\pi n}{L},\;n=1,2,...,
\]
where $L$ determines the variation range of the electron coordinate
normal to the interface: $z\in$\linebreak
$\in\lbrack-L/2,+\infty)$.\,\,The parameters $L$ and $S$ tend to
infinity, so that the problem is considered in the thermodynamic
limit.

\section{Effective Potential\\ of Electron-Electron Interaction}

According to work \cite{KM2006CMP}, the two-dimensional Fourier
transform of the effective potential of electron-electron
interaction with respect to the radius vector $\mathbf{r}_{||}$ can
be expressed in the form\vspace*{-1mm}
 \[
g(q|z_{1} ,z_{2} ) = \nu( q | z_{1} - z_{2} )\, + \]\vspace*{-9mm}
\[
+\, \frac{\beta }{S L^{2} } \!\!\!
       \int\limits_{-L/2}^{+\infty}\!\!\!\! \dd z\!\!\!
       \int\limits_{-L/2}^{+\infty}\!\!\!\! \dd z'  \nu ( q | z_{1}  - z)\,\times\]\vspace*{-7mm}
\begin{equation}\label{ekran}
       \times\, {{\mathfrak M}}( q| z, z') \nu(q | z'- z_{2} ),
 \end{equation}
where\vspace*{-3mm}
\[
{\nu(q|z_{1}-z_{2})=\frac{2\pi e^{2}}{q}\,\mathrm{e}^{-q|z_{1}-z_{2}|}}%
\]
is the two-dimensional Fourier transform of the Coulomb potential, $z_{1}$ and
$z_{2}$ are the coordinates of electrons reckoned normally to the interface,
$\beta$ the inverse thermodynamic temperature, and ${{\mathfrak{M}}%
}(q|z,z^{\prime})$ the \textquotedblleft den\-si\-ty--den\-si\-ty\textquotedblright%
\ correlation function, which is a solution of the Fredholm integral
equation of the second kind \cite{KMK2006ICMP},%
 \[{{\mathfrak M}}(q|z_{1},z_{2})=
  {{\mathfrak M}}_{0}(q|z_{1},z_{2})\,+ \]\vspace*{-6mm}
  \[
  +\,\frac{\beta }{S L^{2} } \!\!\!
  \int\limits_{-L/2}^{+\infty}\!\!\!\! \dd z\!\!\!
  \int\limits_{-L/2}^{+\infty}\!\!\!\! \dd z'
  {{\mathfrak M}}_{0}(q|z_{1},z)\,\times\]\vspace*{-6mm}
  \begin{equation}\label{IntEq}
\times\,\Big(\!\nu(q|z-z')-\overline{\nu}(q|z-z')\!\Big) {{\mathfrak
M}}(q|z',z_{2} ),
 \end{equation}
where ${\overline{\nu}_{k}(q)=G_{k}(q)\nu_{k}(q)}$,
${\nu_{k}(q)=\frac{4\pi e^{2}}{q^{2}+k^{2}}}$ is the
three-dimensional Fourier transform of the Coulomb potential (the
variable $k$ is responsible for the Fourier expansion along the
electron coordinate normal to the interface),\vspace*{-3mm}
\[
{\overline{\nu}(q|z-z^{\prime})=\frac{1}{L}\sum_{k}\mathrm{e}^{\mathrm{i}%
k(z-z^{\prime})}\overline{\nu}_{k}(q)},
\]
${{\mathfrak{M}}}_{0}(q|z_{1},z_{2})$ is the two-particle \textquotedblleft
density--density\textquotedblright\ correlation function in the ideal exchange
approximation \cite{KMK2006ICMP}, and $G_{k}(q)$ is the local-field correction.

In the low-temperature limit, the following expression was obtained in work
\cite{KM2006CMP} for the two-particle correlation function of an electron gas in
the ideal exchange approximation:
 \[
  {\mathfrak M}_0({q}|z_1,z_2)   = \displaystyle\frac{L^2}{\beta}
  \sum\limits_{\alpha_1,\alpha_2}
  \Lambda_{\alpha_1,\alpha_2}({q})\,\times\]\vspace*{-7mm}
\begin{equation}\label{CorrF2}
\times\, \varphi^*_{\alpha_1}(z_1)
  \varphi^{\vphantom{*}}_{\alpha_2}(z_1)
   \varphi^*_{\alpha_2}(z_2)
  \varphi^{\vphantom{*}}_{\alpha_1}(z_2),
 \end{equation}
where\vspace*{-3mm}
\[
      \Lambda_{\alpha_1,\alpha_2}({q})  =  \displaystyle
     \frac{2m}{\hbar^2}\frac{S}{2\pi}
     \frac{\alpha_1^2-\alpha_2^2-q^2}{q^2}\,\times\]\vspace*{-6mm}
\[ \times\, \Bigg[1-
     \sqrt{1-4q^2\frac{p_\mathrm{F}^2-\alpha_1^2}{(\alpha_1^2-\alpha_2^2-q^2)^2}}\,
     \times\]\vspace*{-6mm}
 \begin{equation}\label{Lambda2}
  \times\,    \displaystyle \theta
      \left(\!1-4q^2\frac{p_\mathrm{F}^2-\alpha_1^2}{(\alpha_1^2-\alpha_2^2-q^2)^2}\!\right)\!
   \Bigg]\theta(p_\mathrm{F}-\alpha_1),
 \end{equation}
$p_{\mathrm{F}}=(9\pi/4)^{1/3}/r_{\mathrm{S}}$ is the Fermi
momentum, and $r_{\mathrm{S}}$ the Brueckner parameter in the units
of the Bohr radius~$a_{\mathrm{B}}$.

In work \cite{KM2006CMP}, it was shown that, in some approximations, an
analytical expression can be obtained for the function ${{\mathfrak{M}}}%
_{0}(q|z_{1},z_{2})$,%
    \[{\mathfrak M}_{0}({q}|z,z') = -\displaystyle\frac{SL^2}{\beta}
   \frac{2m}{\hbar^2}\frac{1}{\pi^2}\frac{   \ee^{-q|z-z'|}-\ee^{-q|z+z'|}
   }q\,\times\]
\[ \times\,  \left[
   \dfrac{p_\mathrm{F}\cos(p_\mathrm{F}(z+z'))}{(z+z')^2}
  -\dfrac{p_\mathrm{F}\cos(p_\mathrm{F}(z-z'))}{(z-z')^2}\right.\,+\]\vspace*{-7mm}
\[
+\left. \dfrac{\sin(p_\mathrm{F}(z-z'))}{(z-z')^3}
  -\frac{\sin(p_\mathrm{F}(z+z'))}{(z+z')^3}
   \right]\times\]\vspace*{-7mm}
\begin{equation}\label{M02as}
   \times\,\theta(-z)\theta(-z').
\end{equation}
From this expression, one can see, in particular, that, besides the
terms, which depend on ${(z-z^{\prime})}$ and are characteristic of
uniform systems, there are terms depending on
${(z+z^{\prime})}$.\,\,The reason for their appearance is the
presence of the plane in\-ter\-fa\-ce.\,\,This means that, in our
ideal-exchange approximation for the two-particle \textquotedblleft
den\-si\-ty--den\-si\-ty\textquotedblright\ correlation function,
the image-force effects are already taken into
con\-si\-de\-ra\-ti\-on.\,\,No\-te that the polarization operator in
works \cite{Sidyakin,Bechstedt} is presented as a sum of two
polarization operators for the uniform electron gas: one of them
depends on ${(z-z^{\prime})}$, and the other on
${(z-z^{\prime})}$.\,\,Ho\-we\-ver, formula
(\ref{M02as}) shows that the de\-pen\-den\-ces of ${{\mathfrak{M}}}_{0}(q|z_{1}%
,z_{2})$ on ${(z-z^{\prime})}$ and ${(z+z^{\prime})}$ are not
\mbox{so simple.}

In the following numerical calculations of the two-particle \textquotedblleft
density--density\textquotedblright\ correlation function in the ideal exchange
approximation, expression (\ref{CorrF2}) is used.

\section{Results of Numerical Calculations of Two-Particle \textquotedblleft
Density--Density\textquotedblright\ Correlation Function and Effective
Electron-Electron Interaction Potential}

The two-particle \textquotedblleft density--density\textquotedblright%
\ correlation function of electrons, ${{\mathfrak{M}}}(q|z_{1},z_{2})$, was
numerically calculated according to Eq.~(\ref{IntEq}) and making allowance for
the local-field correction $G_{k}(q)$ taken from the theory of uniform
electron gas in the following forms:

1) the modified Hubbard correction \cite{Gorobchenko}
\begin{equation}
G_{k}(q)=\frac{1}{2}\frac{q^{2}+k^{2}}{q^{2}+k^{2}+\xi p_{\mathrm{F}}^{2}},
\label{GrindEQ__3_}%
\end{equation}
where $\xi$ is a parameter, the values of which are given below; and

2) the Ichimaru correction \cite{Ichimaru}%
 \[
   G_{k} (q)=A Q^{4} +B Q^{2} +C\,+  \]\vspace*{-9mm}
\[+ \left[AQ^4+\left(\!B+\frac83A\right)Q^2-C \right] \times\]\vspace*{-7mm}
\begin{equation}
\times\, \frac{4-Q^2}{4Q}\ln\left|\frac{2+Q}{2-Q}\right|\!,
 \end{equation}

\begin{figure}
\vskip1mm
\includegraphics[width=7.5cm]{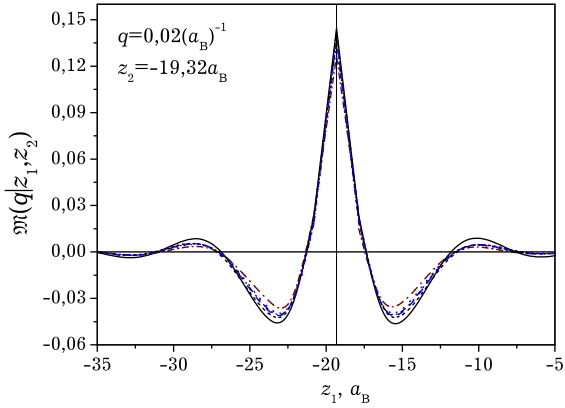}
\vskip-3mm\caption{Dependences of the dimensionless two-particle
``density--density'' correlation function on
the coordinate of electron~1 normal to the interface, whereas the
corresponding coordinate of electron~2 is fixed ($z_{1}= -19.57a_{\mathrm{B}%
}$); $q=0.02a_{\mathrm{B}}^{-1}$  }
\end{figure}

\begin{figure}[h!]
\vskip1mm
\includegraphics[width=7.5cm]{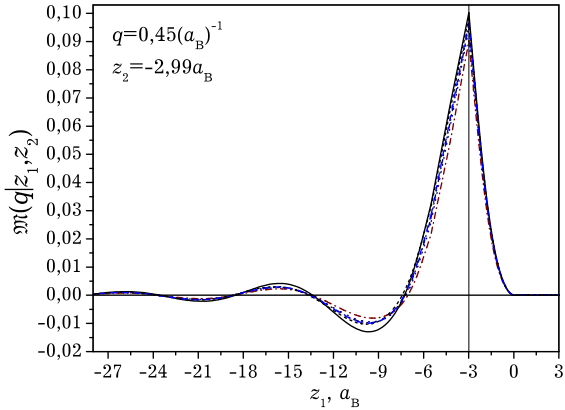}
\vskip-3mm\caption{The same as in Fig.~1, but for
${z_{1}=-2.99a_{\mathrm{B}}}$ and $q=$ $=0.45a_{\mathrm{B}}^{-1}$  }
\end{figure}

\begin{table}[h!]
\vskip3mm \noindent\caption{Legends of the figures
}\vskip3mm\noindent\tabcolsep24.3pt

\noindent{\footnotesize\begin{tabular}{|l|l|}
  \hline
\multicolumn{1}{|c}{\rule{0pt}{5mm}Line type} &
\multicolumn{1}{|c|}{Approximation } \\[2mm]
  \hline \rule{0pt}{5mm}$^{\rule{10mm}{0.8pt}}$ & Animalu   \\
  $^{\rule{1.3mm}{0.8pt}\rule{1.3mm}{0pt}\rule{1.3mm}{0.8pt}\rule{1.3mm}{0pt}\rule{1.3mm}{0.8pt}\rule{1.3mm}{0pt}\rule{1.3mm}{0.8pt}}$ & Geldart and Vosko ($\xi=2$)  \\
  $^{\rule{1.3mm}{0.8pt}\rule{1.3mm}{0pt}\rule{0.3mm}{0.8pt}\rule{1.3mm}{0pt}
    \rule{1.3mm}{0.8pt}\rule{1.3mm}{0pt}\rule{0.3mm}{0.8pt}\rule{1.3mm}{0pt}
    \rule{1.3mm}{0.8pt}}$ & $G_{k}(q)\equiv0$ \\
 $^{\rule{1mm}{0.8pt}\rule{1mm}{0pt}\rule{1mm}{0.8pt}\rule{1mm}{0pt}\rule{1mm}{0.8pt}\rule{1mm}{0pt}\rule{1mm}{0.8pt}\rule{1mm}{0pt}\rule{1mm}{0.8pt}}$ & Hubbard ($\xi=1$) \\
  $^{\rule{0.3mm}{0.8pt}\rule{1mm}{0pt}\rule{0.3mm}{0.8pt}\rule{1mm}{0pt}\rule{0.3mm}{0.8pt}\rule{1mm}{0pt}\rule{0.3mm}{0.8pt}\rule{1mm}{0pt}\rule{0.3mm}{0.8pt}\rule{1mm}{0pt}\rule{0.3mm}{0.8pt}\rule{1mm}{0pt}\rule{0.3mm}{0.8pt}\rule{1mm}{0pt}\rule{0.3mm}{0.8pt}}$ & Sham ($\xi=1+\frac{4}{\pi p_{\mathrm{F}}a_{\mathrm{B}}}$) \\
  $^{\rule{1.3mm}{0.8pt}\rule{1.3mm}{0pt}\rule{0.3mm}{0.8pt}\rule{1.3mm}{0pt}
    \rule{0.3mm}{0.8pt}\rule{1.3mm}{0pt}\rule{1.3mm}{0.8pt}\rule{1.3mm}{0pt}\rule{0.3mm}{0.8pt}\rule{1.3mm}{0pt}
    \rule{0.3mm}{0.8pt}}$ & Animalu ($\xi=1+\frac{2}{\pi p_{\mathrm{F}}a_{\mathrm{B}}}$) \\[2mm]
  \hline
\end{tabular}}{\vspace*{-3mm}}
\end{table}

 \noindent
where ${Q=\sqrt{q^{2}+k^{2}}/p_{\mathrm{F}}}$, and the parameters
$A$, $B$, and $C$ are cumbersome and can be found in work
\cite{Ichimaru}.

All numerical calculations were carried out for potassium ($r_{\mathrm{S}%
}=4.86\,a_{\mathrm{B}}$).

Substituting the numerical solution of Eq.\,(\ref{IntEq}) into
formula (\ref{ekran}), we obtain a two-dimensional Fourier transform
of the effective potential of elect\-ron-elect\-ron
interaction.\,\,Making the inverse Fourier transformation with
respect to the variable $\mathbf{q}$ and taking into account that
the two-di\-men\-sio\-nal Fourier transform of the effective
potential depends only on the absolute value of the vector
$\mathbf{q}$, we obtain the
effective potential of elect\-ron-elect\-ron interaction in the form%
\[
  g(r_{\parallel},z_1,z_2)=\frac1S\sum_{\mathbf{q}}\ee^{\ii \mathbf{q}
  \mathbf{r}_{\parallel}}g(q|z_1,z_2)=\]\vspace*{-8mm}
\[ =\frac1{2\pi}\int\limits_0^\infty\!\!\dd q\, q\,
\mathrm{J}_0(qr_{\parallel})g(q|z_1,z_2),
 \]\vspace*{-4mm}

\noindent where $\mathrm{J}_{0}(x)$ is the zeroth-order cylindrical Bessel
function.

In Figs.\,\,1 to 3, the results of calculations of the two-particle
\textquotedblleft density--density\textquotedblright\ correlation
function of electrons, ${\mathfrak{M}}(q|z_{1},z_{2})$, obtained by
solving the Fredholm integral equation of the second kind
(\ref{IntEq}) making allowance for various local-field corrections
and without them (the random-phase approximation) are
depicted.\,\,If one of the electrons is in the metal depth, this
function is symmetric with respect to the coordinate of a fixed
electron directed normally to the interface plane and has a sharp
peak, when those coordinates of two electrons coincide (see Fig.~1),
i.e.\,\,the electrons correlate with each other and do not feel the
influence of the surface.\,\,If one of the electrons approaches the
interface, the latter starts to affect the two-particle correlation
function of electrons.\,\,The sharp symmetric peak that was observed
in Fig.~1 loses its symmetry and broadens: besides electron
correlations, there emerges an effective repulsion from the
interface (see Fig.~2).\,\,This repulsion results in the following.
When the electron approaches the interface even more, the maximum in
the two-particle correlation function of electrons does not occurs,
when their coordinates coincide, as it was in the metal depth (see
Fig.~1), but is a little shifted to the left from the interface (in
Fig.~3, this maximum is located at about $-2.7a_{\mathrm{B}}$).

\begin{figure}
\vskip1mm
\includegraphics[width=7.5cm]{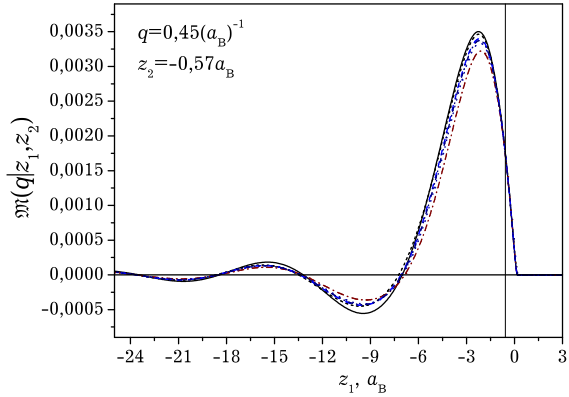}
\vskip-3mm\caption{The same as in Fig.~1, but for
${z_{1}=-0.57a_{\mathrm{B}}}$ and $q= $ $=0.45a_{\mathrm{B}}^{-1}$ }
\end{figure}

\begin{figure}
\vskip1mm
\includegraphics[width=7.5cm]{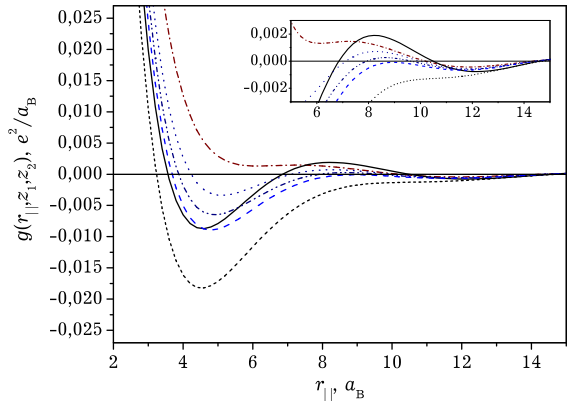}
\vskip-3mm\caption{Dependences of the effective electron-electron
interaction potential on the distance between the electrons along
the interface.\,\,The electron coordinates $z_{1}$and $z_{2}$ normal
to the interface are identical and fixed.\,\,The case
$(z_{1},z_{2})\rightarrow-\infty$.\,\,The notation of curves is the
same as in Fig.~1  }
\end{figure}

In addition, Figs.\,\,1 to 3 demonstrate that various models of the
local-field correction do not change the behavior of the
two-particle correlation function of electrons qualitatively, but do
it quantitatively.\,\,The application of the random-phase
approximation produces the smallest deviations in the two-particle
correlation function of electrons, whereas the Ichimaru correction
leads to the largest ones.\,\,The calculated values of two-particle
correlation function for electrons with the use of other local-field
corrections fall within the interval between the values obtained in
the random-phase approximation and with the use of the Ichimaru
correction.

\begin{figure}
\vskip1mm
\includegraphics[width=7.5cm]{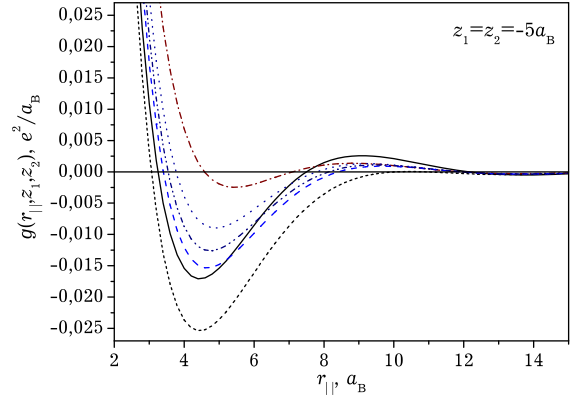}
\vskip-3mm\caption{The same as in Fig.~4, but for
$z_{1}=z_{2}=-5a_{\mathrm{B}}$  }
\end{figure}

\begin{figure}
\vskip3mm
\includegraphics[width=7.5cm]{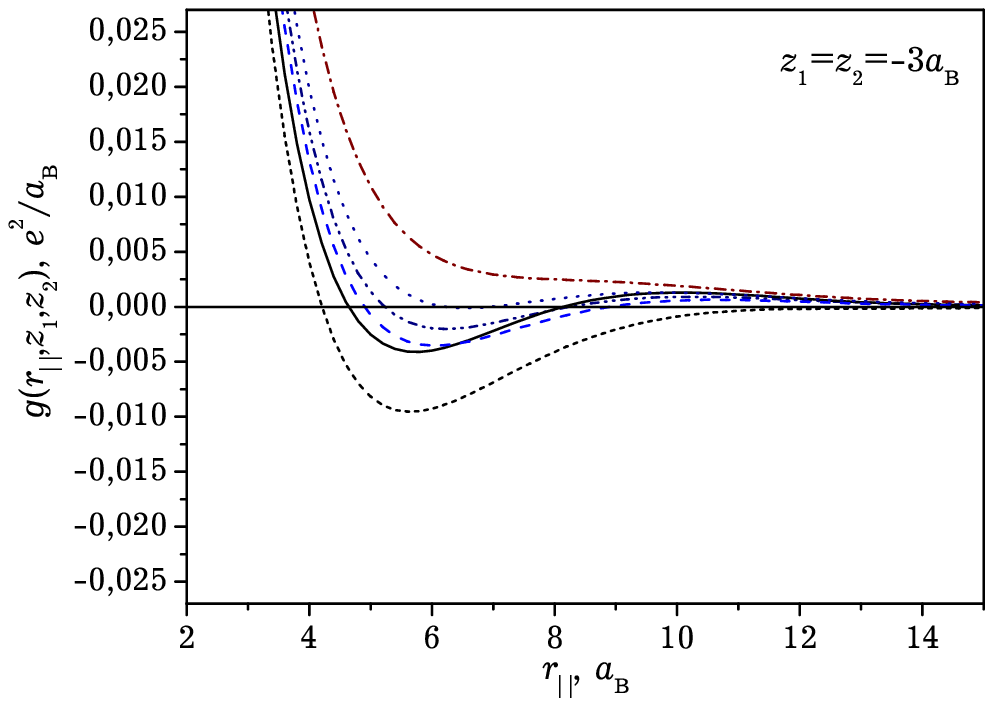}
\vskip-3mm\caption{The same as in Fig.~4, but for
$z_{1}=z_{2}=-3a_{\mathrm{B}}$  }
\end{figure}

\begin{figure}[h!]
\vskip3mm
\includegraphics[width=7.3cm]{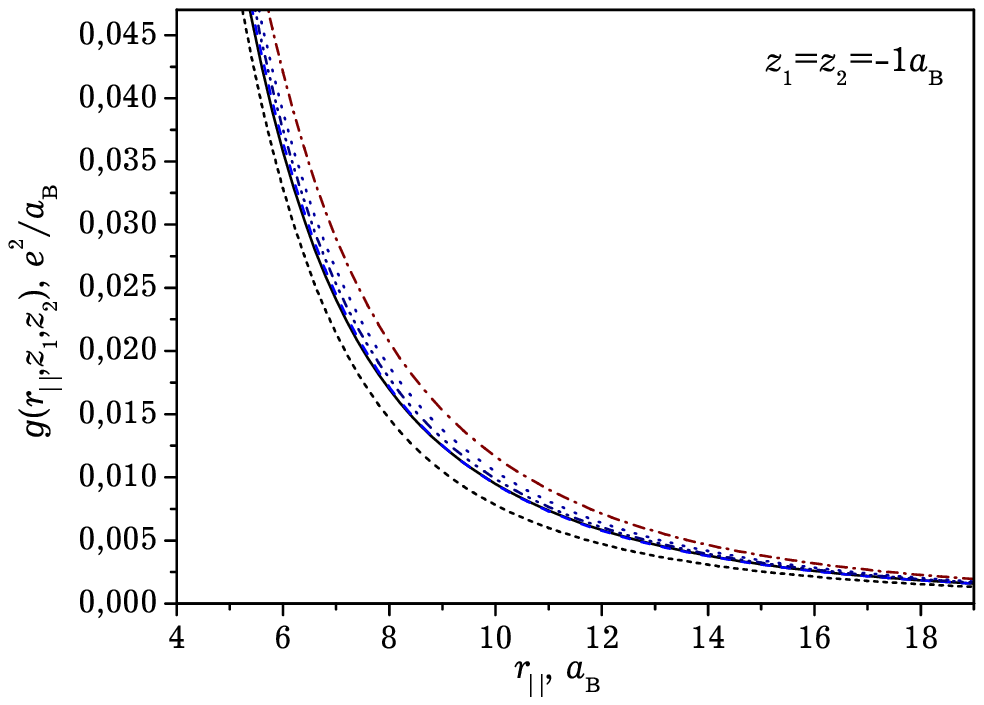}
\vskip-3mm\caption{The same as in Fig.~4, but for
$z_{1}=z_{2}=-a_{\mathrm{B}}$  }
\end{figure}

\begin{figure}
\vskip1mm
\includegraphics[width=7.5cm]{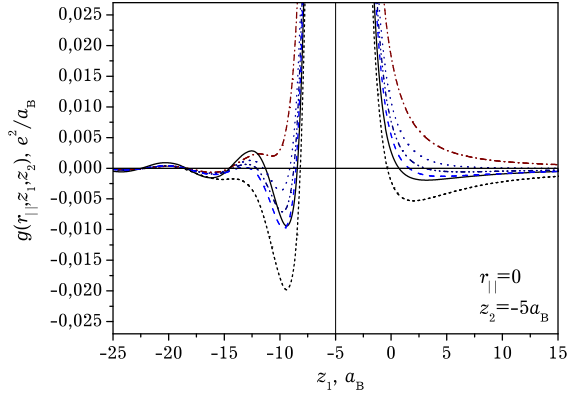}
\vskip-3mm\caption{Dependences of the effective electron-electron
interaction potential on the distance from electron~1 to the
interface.\,\,The normal coordinate of electron~2 is fixed,
${z_{2}=-5a_{\mathrm{B}}}$, $r_{\parallel}=$ $=0$.\,\,The notation
of curves is the same as in Fig.~1  }
\end{figure}

\begin{figure}
\vskip3mm
\includegraphics[width=7.5cm]{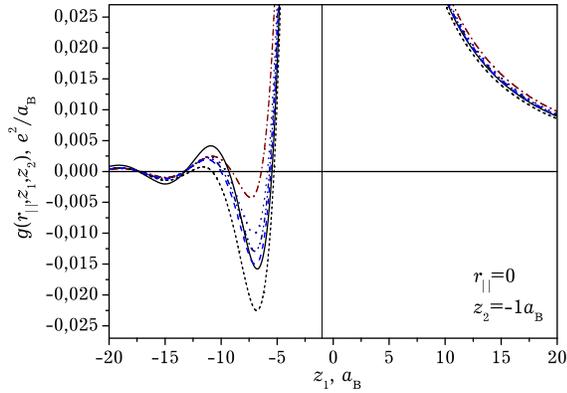}
\vskip-3mm\caption{The same as in Fig.~8, but for
$z_{2}=-a_{\mathrm{B}}$  }
\end{figure}

In Fig.\,\,4 to 9, the results of calculations for the
ef\-fec\-ti\-ve potential of elect\-ron-elect\-ron interaction
$g(r_{\parallel},z_{1},z_{2})$ obtained for various local-field
correction models and without them (the ran\-dom-pha\-se
app\-ro\-xi\-ma\-ti\-on) are shown.\,\,In Fig.~4, the effective
potential of interaction between electrons located in the metal
depth, i.e.\,\,when they do not feel the interface influence, is
exhibited.\,\,From this figure, one can see that making allowance
for the local-field correction brings about the appearance of a
potential well at distances from 4.5 to 5 times $a_{\mathrm{B}}$,
depending on the specific correction model.\,\,The depth of this
potential well also depends on the local-field correction
model.\,\,The deepest potential well corresponds to the Hubbard
correction, and the shallowest one to the Sham model.\,\,Other
examined corrections give rise to intermediate depths of the
potential well.\,\,In addition, in the random-phase approximation
(i.e.\,\,when the local-field correction is not taken into
consideration, the dash-dotted curve), there is no potential well at
indicated distances in the case of a uniform system.\,\,At large
distances, there emerge Friedel oscillations
\cite{Gabov1,Gabov2,Harrison}.\,\,However, in the case of the
effective potential calculated in work \cite{Fratesi}, they are
absent.

In Fig.\,\,5, the dependence of the effective electron-electron
interaction potential on the distance between the electrons is
depicted for the normal coordinates of electrons
${z_{1}=z_{2}=-5a_{\mathrm{B}}}$.\,\,For this distance of electrons
from the interface, a considerable deepening of the potential well
is observed; moreover, it appears even in the random-phase
approximation.\,\,This fact originates from a nonmonotonic behavior
of the electron density distribution $n(z)$ near the interface
\cite{JPS2003:2}; namely, here, the plane layers with electron
concentrations lower and higher than that in the metal depth
alternate.\,\,As a result, the collective effects in the
electron-enriched layers are more pronounced, and the electron
screening is stronger.\,\,In the electron-depleted layers, the
situation is opposite: the screening is weaker, and the repulsion
between electrons becomes stronger (see Fig.~6).\,\,If the electrons
come nearer to the interface, the repulsion between them prevails
(Fig.~7).\,\,As the coordinates of electrons normal to the interface
grow further, the effective potential of electron-electron
interaction tends to the Coulomb potential:\vspace*{-3mm}
\[
\lim\limits_{z_{1},z_{2}\rightarrow\infty}g(r_{\parallel},z_{1},z_{2})=\frac{e^{2}%
}{\sqrt{r_{\parallel}^{2}+(z_{1}-z_{2})^{2}}}.
\]

The same behavior is demonstrated in Figs.~8 and 9.\,\,They exhibit
the dependence of the effective potential of interaction between the
electrons located on the same normal to the interface
($r_{\parallel}=0$) on the normal coordinate of one of the
electrons, regarding the other electron to be fixed.\,\,The presence
of the interface results in a nonsymmetric effective potential of
electron-electron interaction with respect to the electron
coordinate normal to the interface.\,\,There are more electrons to
the left from the fixed one, and the screening is stronger;
therefore, the potential wells and Friedel oscillations are
observed; to the right, the number of electrons is smaller, so that
the potential well is either shallower or disappears.

\section{Conclusions}

To summarize, the two-particle correlation function of electrons and
the effective potential of electron-electron interaction have been
calculated making allowance for various local-field correction
models known in the theory of uniform electron gas.\,\,The behavior
of the two-particle correlation function of electrons depending on
the electron coordinates normal to the interface is studied, as well
as the influence of various local-field correction models on
it.\,\,In particular, it is found that the presence of the interface
gives rise to an additional effective repulsion of the electrons
from the interface, the maximum of the two-particle correlation
function does not take place at the coincidence of electron
coordinates, as occurs in the metal depth, but is a little shifted
from the \mbox{interface.}\looseness=1

The behavior of the effective electron-electron interaction
potential depending on the electron coordinates normal to the
interface and the distance between the electrons along the interface
is also examined, as well as the influence of various local-field
correction models on it.\,\,The results of our calculations
demonstrate that, in the near-surface region of the metal, there are
plane layers, where the behavior of the effective potential of
electron-electron interaction is essentially different.\,\,Namely,
in some layers, the effective potential oscillates and form deep
potential wells, whereas the wells are shallower or even absent in
other layers.\,\,As a result, the additional mechanical stresses
emerge near the metal surface, which can provoke the appearance of
cracks and other defects.

\rezume{%
Б.М.\,Маркович, І.М.\,Задворняк}{ЕФЕКТИВНИЙ ПОТЕНЦІАЛ\\ МІЖЕЛЕКТРОННОЇ ВЗАЄМОДІЇ\\
ДЛЯ НАПІВОБМЕЖЕНОГО ЕЛЕКТРОННОГО ГАЗУ\\ З ВРАХУВАННЯМ ПОПРАВКИ НА
ЛОКАЛЬНЕ ПОЛЕ} {У роботі проведено чисельний розрахунок ефективного
потенціалу міжелектронної взаємодії та двочастинкової кореляційної
функції електронів ``густина--густина'' для напівобмеженого простого
металу з урахуванням поправки на локальне поле.
 Досліджено вплив на них плоскої поверхні поділу та різних форм
 поправки на локальне поле. Показано, що біля поверхні поділу є області з більшою глибиною
потенціальної ями у ефективному потенціалі міжелектронної взаємодії,
ніж в глибині металу.}

\end{document}